\documentclass[pra,twocolumn,titlepage,nofootinbib,amsmath,preprintnumbers]{revtex4-2}%
\usepackage{amsmath}
\usepackage{amsfonts}
\usepackage{amssymb}
\usepackage{graphicx}%
\usepackage{color}

\begin{document}
\title{Laboratory constraint on the electric charge of the neutron and the neutrino}
\author{Savely~G.~Karshenboim}
\email{savely.karshenboim@mpq.mpg.de}
\affiliation{Ludwig-Maximilians-Universit{\"a}t, Fakult{\"a}t f\"ur Physik, 80799 M\"unchen, Germany}
\affiliation{Max-Planck-Institut f\"ur Quantenoptik, Garching, 85748, Germany}


\begin{abstract}
We revisit constraints on the value of the electric charge of the neutron and the neutrinos as well as on the electric-charge proton-electron difference $e_p+e_e$. We consider phenomenological constraints based on laboratory study of the electrical neutrality of would-be neutral subatomic, atomic, and molecular species under assumption of the conservation of the electric charge in the $\beta$ decay, that relates the values of $e_p+e_e, e_n, e_\nu$. Some of constraints published previously utilized an additional assumption $e_\nu=0$, which we do not.
We dismiss a cosmological constraint at the level of $10^{-35}\,e$ utilized by Particle Data Group (PDG) in their {\em Review of particle properties\/} \cite{rpp2022} as a controversial one which makes the laboratory constraints on $e_\nu$ dominant.

The phenomenological constraints from the available data of laboratory experiments are obtained as $e_p+e_e=(0.2\pm2.6)\times10^{-21}\,e$, $e_n=(-0.4\pm1.1)\times10^{-21}\,e$, and $e_\nu=(0.6\pm3.2)\times10^{-21}\,e$. The ones on $e_p+e_e$ and $e_n$ are at the same level as the related constraints of PDG but somewhat different because of releasing the value of $e_\nu$. Our $e_\nu$ constraint is several orders of magnitude weaker than the controversial cosmological result dominated in the PDG constraint, but several orders of magnitude stronger than the other individual $e_\nu$ constraints considered by PDG.

We also consider consistency of the phenomenological constraints and the Standard model (SM). The SM ignores the mass term of the neutrinos and cannot describe the neutrino oscillations which makes it not a complete theory but a part of it.
We demonstrate that the condition of the cancellation of the triangle anomalies within the complete theory does not disagree with the phenomenological constraints since different extensions of the SM may produce different additional contributions to the anomalies. A choice of the extension fixes the way how those contributions are organized. In particular, we consider a minimal extension of the SM, where leptons ($\nu,e$) are treated the same ways as quarks, which sets $e_p+e_e=0$ and allows for numerical strengthening the constraint on $e_n$ and $e_\nu$, which is $e_n=-e_\nu=(-0.4\pm1.0)\times10^{-21}\,e$.
\end{abstract}

\today
\maketitle

\section{Introduction\label{s:intro}}

The value of the electric charge of various fundamental and compound particles is an important quantity that has been of interest for a while. It is experimentally observed that while the subatomic particles are charged, the atomic bulk matter {\em seems\/} neutral. In many evaluations of the values of fundamental constants it is expected that the charge of all the particles is an integer multiple of a certain elementary charge (see, e.g, \cite{codata2018}), i.e., that the electric charge is quantized.
Usually it is considered that the would-be neutral objects, such as the hydrogen atom, the neutron, and the neutrino, {\em are\/} neutral by default mostly for traditional reasons and for a reason that small numbers seldom appear in Nature. It is hard to expect for a nonzero value such as $|e_p+e_e|\ll e_p$ or $|e_n|\ll e_p$. It is also of a great simplification to assume for the education, for [precision] applications, and tests of various electrodynamic effects on atomic and subatomic particles that the charge is quantized and therefore there is only one parameter, that determines the strength of the electromagnetic interaction in quantum electrodynamics (QED), namely, the fine structure constants $\alpha=e^2/4\pi\hbar c$. While that simplifies the interpretation of precision applications of QED theory and its comparison to experiment (see, e.g., \cite{codata2018,QED}), that is not a requirement of the QED by itself (see, e.g., \cite{weinberg,peskin}).

Experimentally, the neutrality of the would-be neutral objects is not obvious. E.g., the force between two massive bodies of the laboratory scale, such as ones with a kilogram-range mass, follows the Coulomb/Newton-type $1/r^2$ law and might be a combination of the Newton force between the masses and the Coulomb force of the charges of the objects.
We can distinguish between those forces only with a certain accuracy which opens the door for a possible [very small] residual charge of the would-be neutral bulk matter. The {\em exact\/} neutrality is a matter of the interpretation and theoretical constructs rather than a `direct' experimental fact. (We mean here the current situation, while the expectation of the exact neutrality by default has taken place since the time when we understood Nature essentially less than now.)

In principle, a possible presence of various kinds of the residual charge of bulk matter is sometimes a source of systematic effects in high-precision or high-sensitivity experiments ranging from laboratory macroscopic measurements
on ultraweak forces to high-resolution spectroscopic experiments in atomic physics (see, e.g., \cite{charge2}).

Considering the neutrality of a bulk matter we have to distinguish two options. One can study a piece of bulk matter as it is. It is technically easy to establish the neutrality of a conductor by grounding it. However, the overall neutrality in the case of $e_p+e_e\neq 0$ or $e_n\neq 0$ can be achieved by having a certain deficit or excess of the electrons over the protons. Or {\em vice versa\/} a residual charge of a piece of a bulk matter can be provided in case of $e_p+e_e= 0$ or $e_n= 0$ by a certain deficit or excess of the electrons. The insulator materials may have a residual charge. However, in both cases, speaking about the constraints we are rather interested in the bulk matter with the same number of protons and electrons, so its charge or the neutrality would have direct consequences for fundamental values such as $e_p+e_e$ and $e_n$. (We refer to $e_p+e_e$ as to the proton-electron charge difference, since we deal with the difference of the absolute values of the charges. Commonly, term `electron charge' is used in two opposite meanings, namely as the absolute value of the charge and as the [negative] charge by itself. In this paper $e_p$ is the positive charge of the proton and $e_e$ is the negative one of the electron.)

Anyway, the measurement of the value of the electric charge of would-be neutral objects, such as the neutron and the neutrino, and the production of the related constraints has been a subject of a certain interest (see, e.g., \cite{rpp2022}). (We indeed consider the charge of the neutrinos being the same for all three generations.) Various phenomenological constraints and theoretical speculations have been made. A non-zero value of either of two mentioned quantities used to be considered as a violation of a standard-physics picture and a demonstration of new physics. The inconsistency of their non-zero value with the standard picture became obvious with the creation of the Standard Model. The latter was created as a model of the electroweak and strong interactions with $\nu_L$ and without $\nu_R$ (see, e.g., \cite{weinberg,peskin}) in a period, when the neutrinos were assumed massless, and {\em then\/} such a consideration seemed complete.

The discovery of the neutrino oscillations (see, e.g., \cite{zuber}) has changed the situation. The name `Standard model' (SM) is widely in use in a broad context, however the SM (in its narrow sense) is incomplete as a theory that in particular covers the electroweak interaction of leptons, at least because of the presence of the neutrino oscillations, a description of which requires certain neutrino mass terms. The term `SM' is often understood as a kind of a base theory of experimentally established phenomena partly because many text books ignore the oscillations in the most of the text and partly because there is no single consensus model of theory of the electroweak and strong interactions with presence of $\nu_R$ and/or the neutrino mass term. However, while using it, it is not always clear whether one considers the SM as a complete theory (which is incorrect in the exact sense, but may be a good approximation for many phenomena), or as a well established part of the [complete, but unknown] Lagrangian of the fundamental particles, including all $\nu$'s details, and their electroweak and strong interactions. Parameterizations of natural phenomena within the SM can be only approximate and in context of the small possible values of the charge of the would-be neutral objects may be not applicable.

There is no consensus on how the mass of the neutrino should be accommodated and combined with the SM. The first consequence is the possibility for a nonzero value of the neutrino charge $e_\nu\neq 0$. A massive spin-1/2 particle, does not matter whether it is [very] light, is always suitable to carry an electric charge. Without a consensus on the theory that describes the results of the SM with $m_\nu\neq 0$, we may set only phenomenological constraints. We have empiric constraints of the values of $e_n, e_\nu, e_e+e_p$ from study with free particles, such as the neutrino and the neutron, and with the atomic and molecular objects, that consist of electrons, protons, and neutrons as, e.g., reviewed in \cite{rpp2022}. The only additional theoretical relation between them, that we could use for the empiric constraints, is
the charge conservation for the $\beta$ decay [of the neutron] and related phenomena.

When long time ago it was decided by default that $e_p+e_e=0$ and $e_n=0$, there was no reason for that, which would follow from the requirement of the contemporary theory.

Once the massless neutrino was introduced, it was suggested $e_\nu=0$ and that was for a reason, which was a consequence of the massless character of $\nu$. That suggestion was both convenient and natural.
It is indeed reasonable not to introduce a small charge of the neutrino without a need, since that would create theoretical problems.
The conservation of the electrical charge in the $\beta$ decay led then to the expectation of
$e_p+e_e=e_n$.

The next stage was the SM with still expected then massless neutrinos. The condition on the cancellation of the so-called triangle anomaly sets constraint $e_p+e_e=e_n=0$. (The condition of the cancellation is usually expressed in the terms of the quark charges \cite{weinberg}, that are related to the proton's and neutron's ones through the simple linear combination. The charges of the proton and the neutron are the base of atomic physics and therefore they are the ones that measured accurately.) That was the only period when the suggestion on zero values of $e_p+e_e$, $e_n$, and $e_\nu$ was based on the self-consistency of a theory (the SM), a theory that had been then confirmed by the experiment and was a kind of a complete one for all the phenomena with electroweak and strong interaction of the fundamental particles including the [then-massless] neutrinos [with only left components].

Later on, the neutrino oscillations were discovered (see, e.g., \cite{zuber}) and it became clear that a certain neutrino mass term was missing in the SM Lagrangian and should be introduced. The oscillations by themselves through the need for the mass term released the condition on $e_\nu$. Its non-zero value becomes possible. (Again, it is indeed reasonable not to introduce a small charge of the neutrino without a need, but one should not confuse the statement that of no-need for the neutrino charge with the statement that the [small] neutrino charge is inconsistent with the contemporary overall theoretical construction, that includes the SM (without $\nu_R$ and $m_\nu$) together with an empiric description of the neutrino oscillations that are not included into SM.)

The conservation of the charge for the $\beta$ decay is therefore to be read as $e_p+e_e=e_n+e_\nu$, while the condition for the cancellation of the triangle anomaly is released and becomes uncertain because the complete theory may allows for $\nu_R$ contributions to the triangle anomaly, but their value depends on how the introduction of new terms in the neutrino sector is administrated, which makes it dependent on the extension of the SM model.

Still, there is a theory that includes no additional mechanisms for the masses (like Majorana mass), no additional field (like additional Higgs fields), no additional symmetries (like a symmetry that may set a zero mass for the lightest neutrino and therefore `protect' it from the radiative corrections), and no additional symmetry-breaking term (like the Majorana-mass term). Such a theory treats the quarks and leptons in the same way, while the neutrino mass is a Dirac-type one that originates from the same vacuum average $v$ of the Higgs field as the masses of other particles that are massless prior the symmetry breaking.

We refer to such a model as to the [minimally] extended Standard model. Its symmetry is ${\rm SU}_C(3)\times{\rm SU}_L(2)\times{\rm U}_Y(1)$ that is spontaneously broken by the Higgs mechanism with the minimal Higgs sector (cf. \cite{veltman,jegerlehner}, \cite{weinberg}), while the introduction of an additional ${\rm SU}_L(2)$ singlet $\nu_R$ to the SM offers an additional parameter for the interaction, which is the ${\rm U}_Y(1)$ hypercharge of $\nu_R$. Since the latter is the singlet its hypercharge is proportional to its electric charge.

The values of the charges of the fundamental particles ($u, d, e, \nu_e$---we are focussed on the first generation since the charges for the same members of a generation, say, $\nu$'s, are to be the same through the generations) are fixed within the SM (as a complete theory) by the condition of the cancellation of the triangle anomaly and the condition $e_\nu=0$. Within the extended SM the condition on $e_\nu$ is invalid, while the other cancellation conditions are modified, but maintained.
As we see from the ${\rm SU}_L(2)$ quark doublet $(u_L, d_L)$ it is not necessary that one component of the doublet is neutral. The lack of condition $e_\nu=0$ as an {\em a priori\/} requirement (that would be needed for a massless neutrino) adds one more parameter but does not change the number of the equations for the cancellation of the triangle anomaly (cf. \cite{weinberg}). As a result we can introduce an arbitrary value of $e_\nu$ and nevertheless find a solution for the charges of fundamental particles
that satisfy the condition for the cancellation of the triangle anomalies (see, e.g., \cite{veltman,anomaly1,anomaly3,anomaly4}).

Either we should consider the condition of the cancellation of the triangle anomalies as uncertain for the moment (as far as the extension of the SM is not specified) or modified (once the extension is specified). In either case as a result non-zero values of some would-be neutral charges become possible and consistent with the SM as a {\em part\/} of the complete theory. That opens a new opportunity to look for non-vanishing $e_n$ and $e_\nu$ not as for a part of new physics, but as for a part of establishment on how the SM part of the Lagrangian with the Pontecorvo-Maki-Nakagawa-Sakata neutrino-mixing matrix and the neutrino mass term are to be combined.

The paper is organized as the following. We first overview (see Sect.~\ref{s:pheno}) the existing constraints on $e_n, e_p+e_e$, and $e_\nu$ (mostly following \cite{rpp2022}) and the involved assumptions. We note that even for laboratory phenomenological constraints it often used to be expected that $e_\nu=0$ (see, e.g., \cite{bressi,rpp2022}; cf. \cite{hughes,das,review1}). The laboratory constraints on the charge of would-be neutral atomic substance, that consists of protons, neutrons, and electrons, are sensitive to $e_\nu$ \cite{hughes,das,review1} once we assume the charge conservation in the neutron $\beta$ decay.

We correct phenomenological laboratory constraints for possible $e_\nu\neq0$ and, combining results on the charge of different would-be neutral molecular, atomic and subatomic particles \cite{bressi,baumann,hughes}, derive a constraint on $e_n$ and $e_p+e_e$ (cf. \cite{bressi,rpp2022}), and, assuming the charge conservation in the $\beta$ decay, deduce from them a constraint on $e_\nu$. The results (in units of $e$) are at the level of a part in $10^{21}$. We also study the reliability of constraints on $e_\nu$ that claimed to be stronger than the derived here laboratory constraint.

We note that the laboratory constraints on $e_\nu$ derived here are stronger by several orders of magnitude than other constraints available in literature (see, e.g., \cite{rpp2022}; cf. \cite{hughes,das,review1,review2}) but one \cite{caprini}. The mentioned exceptional constraint \cite{caprini}, that dominates in the evaluation in \cite{rpp2022}, is based on the assumptions that do not make it plausible.

The problem of [a possible nonzero value of] the neutrino charge is also known under a name of `the quantization of the electric charge' and the related conditions were discussed, e.g., \cite{anomaly1,newenu,anomaly2}. The quantization means that we can express all the charges in terms of integer multiples of one of them, say, $e_e$. The quantization of the charge requires $e_\nu=0$. It may be desired under certain ideas of development of the unification schemes, but it is not required for the contemporary theory.

Meantime, quantum electrodynamics (QED) does not require any quantization of the electric charge by itself, however, its `standard' default version is the one with $\alpha=e^2/(4\pi\hbar c)$ and $e=e_p=-e_e$, while $e_n=0$ and therefore the charge of a nucleus is an integer multiple of the proton charge. The original default construction suggested the presence of the `quantum' of the charge, however, later on with the introduction of quarks the `quantum' became three times smaller. This change in the default QED picture has not been considered as any change in QED, which is an important example for distinguishing a theory and its default version---to a certain extend we have to decide whether the SM is a theory with a certain symmetry, broken with a certain mechanism, with certain multiplets (per a generation), a certain number of generations, certain conditions related to the triangle anomalies, and flexible values of the hypercharge of the members of the multiplets or the SM should have exact prearranged values of all the hypercharges.

Another related theoretical concept is the `uniqueness of the SM' \cite{veltman,jegerlehner}, which is a more advanced framework than the quantization of the charge. Since it assumes the SM with the originally chosen values of the charge of the fundamental particles, in particular it technically relies on condition $e_\nu=0$ (see below) and it also concerns about the Higgs mechanism with the minimal Higgs sector which narrows the scope of the extensions of the SM.

The Standard model is a model that covers all the [known] fundamental particles (but $\nu_R$) and contains all the terms, kinematic and of interactions, related to them (except the neutrino mass term if it is with the Majorana mass). The values of the electric charge of the involved particles are determined through their hypercharge related to the original ${\rm U}_Y(1)$ symmetry.

One can understand the SM in two ways. One is a model with fixed values of the hypercharge of the fundamental particles. In such a case one has to check whether the chosen values allow for a self-consistent theory. The other option is to consider a model with free values of the hypercharge that are to be {\em found\/} under condition of the self-consistency of the theory and its agreement with the experiment. (The latter has a limited accuracy that is not sufficient to exclude [very] small values of any quantities, but is sufficient to chose between several discrete solutions if necessary.) Considering the second option, it is important whether the SM is a unique theory with a certain symmetry group etc., or one has different options to choose its parameters, such as different $e_\nu$. If the theory is not unique we cannot experimentally distinguish a theory with $e_\nu=0$ and with a sufficiently small value of $e_\nu\neq0$. We can only constraint the value of $|e_\nu|\ll e_p$ experimentally.

Once we consider the SM as a complete model, the criterium of its self-consistency is the cancellation of the triangle anomalies, which is well defined, and the results of both approaches agree \cite{veltman,anomaly1,anomaly3,anomaly4} (see also \cite{weinberg}). However, we do know that the SM {\em does not\/} describe the reality completely. One has to add some neutrino terms and we do not really know how to do that. They may contribute to the anomalies which makes the condition of the cancellation ill-defined.

All the phenomenological constraints discussed in this paper are consistent with the SM (as a part of the complete theory).

Indeed, once an extension of the SM is introduced explicitly it may set additional relations on the charges. E.g., suggesting that $e_\nu=0$ we arrive (within the SM) at a trivial result with $e_p+e_e=e_n=0$ unless we suggest additional terms beyond the SM that contributes to the cancellation of the triangle anomalies.

However, $e_\nu=0$ is not the only option
and in Sect.~\ref{s:esm} we consider the constraints within the extended SM with $e_\nu$ as a free parameter. The charges of the proton, the neutron, the electron, and the neutrino are related and we derive new constraints on $e_n$ and $e_\nu$. The constraints within the extended SM are at the same level as the phenomenological ones, but somewhat stronger. The only exception is the value of $e_p+e_e$ that may be non-vanishing for the phenomenological constraints, while within the extended SM we find $e_p+e_e=0$.

Still, we specifically consider the model-dependence of the condition $e_\nu=0$ and its phenomenological and theoretical consequences in Sect.~\ref{s:enu}. As concerning the latter, we note that the particles of the fundamental presentation (spin 1/2) contributes to the triangle anomaly. The ones of the adjoint representation (the $W$ bosons) do not. The complete result for the triangle anomaly should be zero. Otherwise, we would not have a selfconsistent theory. The {\em complete\/} result might in principle include the contributions beyond those of known quarks and leptons (including the neutrinos). Once we set $e_\nu=0$ we exclude the possible neutrino contribution to the triangle anomaly, while suggesting $e_n\neq0$, we are to consider an alternative to the SM. As a result we have no constraint from the triangle anomaly on $e_\nu, e_n$, but a condition on the existence of spin 1/2 BSM particles that are to participate in electroweak interactions.

The conclusion section of the paper summarize the phenomenological and theoretical constraints on $e_p+e_e, e_n, e_\nu$

Through out the paper we define the elementary charge as $e=-e_e$, and it may be related to the proton charge exactly (as $e_p=e$) or approximately ($|e_p+e_e|\ll e$) depending on the framework that may set $e_p+e_e=0$ exactly or release the relation. Since the paper is devoted to the constraints we express our results in terms of $e_p$ and $e_n$, accessible experimentally through atomic physics etc., rather than in terms of the quark charges $e_u, e_d$, which are indeed simply related.

\section{A brief overview of the key data and a phenomenological constraint\label{s:pheno}}

Below we overview the constrains mostly listed in \cite{rpp2022} that we consider as a standard reference in the field.
{\em Review of particle properties\/} \cite{rpp2022} contains constraints on $e_p+e_e$, $e_n$, and $e_\nu$ in separate sections.
(The related values are not brand fresh, being reproduced from previous {\em Reviews of particle properties\/}. We give here the reference only to the most recent {\em Review\/}.) The constraints, considered in separate sections of \cite{rpp2022}, are not entirely consistent (see below).
We are mostly interested in laboratory constraints. The latter deal with the neutrality of certain atomic or subatomic particles. Generically the {\em technical\/} result of such a constraint can be presented in the form (cf. \cite{das,review1})
\begin{equation}\label{eq:lab:con}
\left(A-Z\right) e_n + Z\left(e_p+e_e\right) = A\, R\,,
\end{equation}
where $A\, R$ is the constraint on the overall charge of a would-be neutral atomic or molecular system (or the neutron) with $Z$ protons and $A-Z$ neutrons.
The charges $e_i$ are the ones of the related particles and $R$ is the excessive charge normalized per a single nucleon. Some of the laboratory results are collected in Table~\ref{tab:q:data}.

\begin{table}[thbp]
\begin{center}
\begin{tabular}{ccccc}
\hline
Particle   & $A$ & $Z$ & $R$ & Ref. \\
\hline
K & 39 & 19 &  $(22\pm20)\times10^{-21}\,e$ & \cite{hughes}\\
Cs & 133 & 55 & $(12\pm5)\times10^{-21}\,e$ & \cite{hughes}\\
$n$ & 1 & 0 & $(-0.4\pm1.1)\times10^{-21}\,e$ & \cite{baumann}\\
SF$_6$ &146&70& $~(-0.1\pm1.1)\times10^{-21}\,e~$ & \cite{bressi}\\
\hline
\end{tabular}
\end{center}
\caption{Experimental constraints on the electrical neutrality of subatomic \cite{baumann}, atomic \cite{hughes}, and molecular \cite{bressi} particles, parameterized following (\ref{eq:lab:con}).}
\label{tab:q:data}       
\end{table}

\begin{figure}[thbp]
\begin{center}
\resizebox{1.0\columnwidth}{!}{\includegraphics{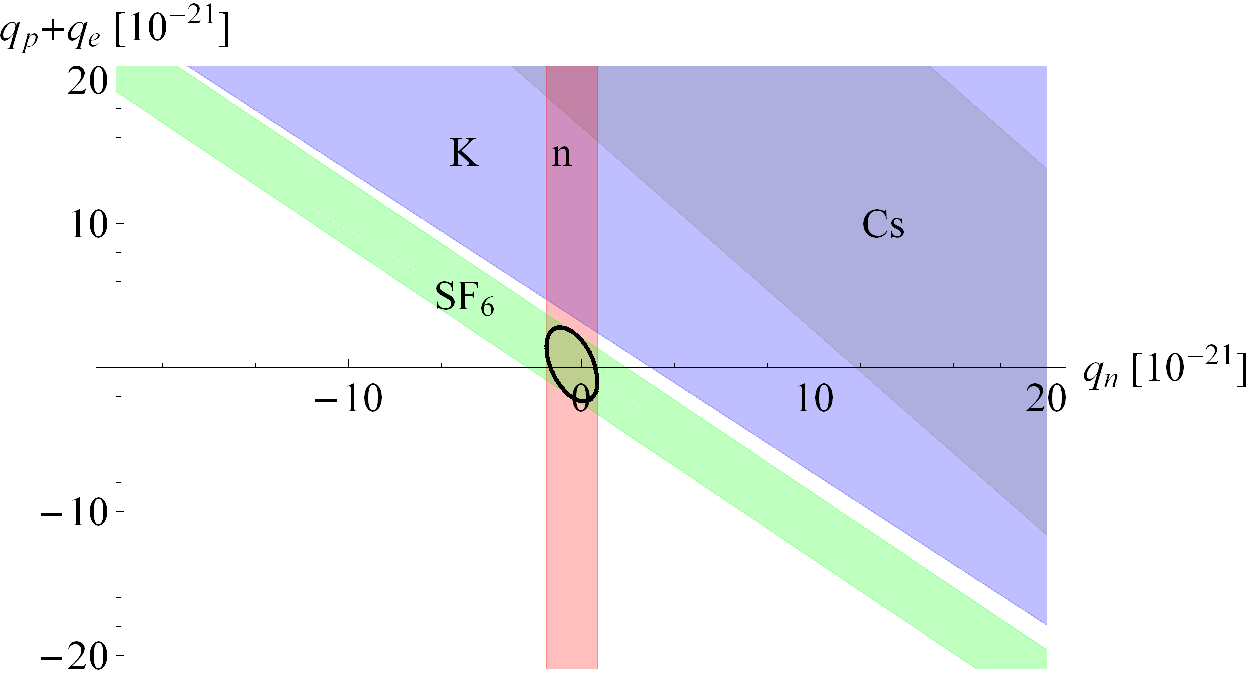}}
\end{center}
\vspace{-3.0mm}
\caption{Phenomenological constraint on the proton-electron charge difference vs. the neutron electric charge from experiments on the electrical neutrality of K and Cs \cite{hughes}, $n$ \cite{baumann}, and SF$_6$ \cite{bressi} (see Table~\ref{tab:q:data} for detail). Our phenomenological average values (derived from the data on $n$ and SF$_6$) (see (\ref{eq:phen:m})) are correlated and the overall constraint is present as the standard ellipse. Here: $q_i=e_i/e$.
}
\label{f:charge:pen}       
\vspace{-4.0mm}
\end{figure}

The results, presented in the table, are `technical' results of the papers and they are not always given explicitly in the related publications, but can be restored from those. The aim of the related publications is usually to present the constraints related to more fundamental quantities, such as $e_p+e_e$ or $e_n$, and a constraint on the charge of the atomic or molecular species really studied experimentally may be not explicitly presented. Moreover, the published constraints, that are rather extracted than directly measured, may be resulted from additional unnecessary assumptions.

To proceed with a constraint on the fundamental quantities one has to make suggestions and, possibly, combine several
technical results. The situation
in general can be described as the following. The phenomenological constraints are either based on specific models being model-dependent or on certain assumptions possibly making them model-dependent. There are two important assumptions applied. One is the charge conservation in the $\beta$ decay of the neutron and the other is the zero value of the neutrino charge.
We consider the constraints derived under assumption of 
the charge conservation as model independent ones since the violation of the charge conservation should have many consequences and, in particular, the violation makes the quantum electrodynamics inconsistent. On contrary, the suggestion of the chargeless status of the neutrino we consider as model dependent (see below). 

For the interpretation of the most accurate experiment on the atomic and molecular particles, which is on sulfur hexafluoride ${\rm SF}_6$ \cite{bressi}, for the interpretation of the molecular constraint in terms the neutron one the charge conservation was applied with additionally assuming $e_\nu=0$. The overall relation between the charges of subatomic particles took the form
\begin{equation}\label{eq:beta:pen}
e_n=e_p+e_e\;.
\end{equation}
Often the conservation in the form of (\ref{eq:beta:pen}) is referred to just as the charge-conservation condition possibly confusing readers (cf. citation of \cite{bressi} in the proton section of \cite{rpp2022}). If one applies the conservation in the form of (\ref{eq:beta:pen}), the technical constraints in (\ref{eq:lab:con}) would turn to
\begin{equation}\label{eq:lab:con:0}
e_n = R\,.
\end{equation}
In particular, the constraint on $e_n$ given by PDG \cite{rpp2022} (see below) is based on constraints from \cite{baumann} and \cite{bressi} under the assumption $e_n=e_p+e_e$, i.e., it is the weighted average of the related $R$'s.

Interpreting an earlier experiment (with K and Cs \cite{hughes}), the charge conservation in the $\beta$ decay was given there in its complete form
\begin{equation}\label{eq:beta:pennu}
e_n+e_\nu =e_p+e_e\;.
\end{equation}
It was also explicitly mentioned in \cite{hughes} that one may {\em additionally\/} consider $e_\nu=0$, which would numerically strengthen the constraint on $e_n$ etc.

Below in this section we first overview the data and next derive a phenomenological constraint.

We do not see any {\em a priori\/} reason to set $e_\nu=0$ and therefore we have to restore the complete form of the constraint on SF$_6$ from \cite{bressi} in the form of (\ref{eq:lab:con}). (Because of the importance of the suggestion $e_\nu=0$ we consider it and its phenomenological and theoretical consequences in detail in Sect.~\ref{s:enu}.)
As shown in Table~\ref{tab:q:data} we have data with different combinations of $A$ and $Z$. That allows us by combining at least two results to reach constraints on $e_p+e_e$ and $e_n$ and, consequently, on $e_\nu$ suggesting the charge conservation as in given (\ref{eq:beta:pennu}).

Once we obtain a laboratory constraint on $e_\nu$, the best of which are at the level of parts in $10^{21}$ (in units of $e$), we have to compare it with available direct constraints on this value. The direct constraints come from cosmology, astrophysics, and reactor physics \cite{rpp2022}. In particular, there is a very strong model-dependent constraint on $e_\nu$ (see, e.g., \cite{caprini}) that we do not accept (see below).

We are interested in rather model-independent constraints on $e_\nu$, that satisfy the electric-charge conservation, and are comparable with the level of the mentioned laboratory constraints. Speaking about the charge conservation as a requirement in context of cosmological considerations we mean the conservation at the current cosmological epoch. As we see below there is no phenomenological constraints on $e_\nu$, that are both more or less model-independent and comparable with the laboratory ones in the strength.

Let's start our overview of the results of an experiment on the neutrality of SF$_6$ molecules described in \cite{bressi}. The constraint in their paper
\[
e_n=(-0.1\pm1.1)\times10^{-21}\,e
\]
was obtained under the assumed charge conservation in form (\ref{eq:beta:pen}), i.e., under the assumption of the charge conservation and $e_\nu=0$. We present the technical result of the paper in a more appropriate form (\ref{eq:lab:con}), suitable for $e_\nu\neq0$, with the parameters listed in Table~\ref{tab:q:data}. Without the suggestion $e_\nu=0$ no constraint on $e_n$ or $e_p+e_e$ from paper \cite{bressi} by itself is possible. To reach such a constraint one has to combine at least constraints on two particles with different values of $Z, A$.

A direct experiment on the neutrality of neutrons \cite{baumann} delivers the result
\[
e_n=(-0.4\pm1.1)\times10^{-21}\,e
\]
that does not involve any additional interpretation or corrections.

Those two mentioned results (on SF$_6$ and $n$) are the ones on base of which the PDG average for the neutron \cite{rpp2022}
\begin{equation}\label{eq:en:rpp}
e_n=(-0.2\pm0.8)\times10^{-21}\,e
\end{equation}
is found. As mentioned, they applied the results on SF$_6$ from \cite{bressi} under a simplified form of the conservation of the electric charge as in (\ref{eq:beta:pen}) and therefore the average of two experiments should be corrected (see below). A related PDG result on $e_p+e_e$ is given in another section of \cite{rpp2022} based on the SF$_6$ alone while assuming a simplified version of the charge conservation in form (\ref{eq:beta:pen}).

The electrical neutrality of two atomic species, K and Cs, has been investigated in \cite{hughes}, where they used a notation with $q_X$ for the charge of particle $X$ and $q_e$ as the unit. It is not very clear what is the sign of $q_e$ accepted in \cite{hughes}. E.g., the use of various inequalities and referring to $q_e$ as to a `magnitude' means that $q_e$ is rather positive, while Eq. (2) in their paper is more consistent with negative $q_e$. Moreover, definition $\delta q =q_e-q_p$ in p. 146 of \cite{hughes} and expression for the atomic charge $q(Z)=Z\delta q + (A-Z) q_n$ in p. 150 indicate a possible typo. In the meantime, the expression was used for a subsequent evaluation. We only partly reproduce their results on the charges of fundamental particles (under assumption, that $q_e$ is positive). The results, that are reproduced, are reproduced approximately since the experimental constraints on K and Cs are correlated, but relatively weakly \cite{hughes}, without the correlation coefficient given. Thus we should neglect the correlation in our evaluation. We present the details related to the results of the experiments on the neutrality of K and Cs \cite{hughes} in Table~\ref{tab:q:data} and further on in several plots (see, e.g., Fig.~\ref{f:charge:pen}) assuming positivity of $q_e$.

The conservation of the electric charge (\ref{eq:beta:pennu}) allows one to express the constraints on the charge of fundamental particles, combining results on the neutrality of K and Cs. The result reads \cite{hughes}
\begin{eqnarray}\label{eq:hughes}
e_p+e_e&=&(0.9\pm2.0)\times10^{-19}\,e\nonumber\\
e_n&=&(0.4\pm1.5)\times10^{-19}\,e\nonumber\\
e_{\nu_e}&=&(0.5\pm3.5)\times10^{-19}\,e\,.
\end{eqnarray}
We give here the results for $e_p+e_e$ and $e_n$ as published in the cited paper (see Eq. (5) there), assuming expression $q(Z)=Z\delta q +N q_n$ is correct while definition $\delta q =q_e-q_p$ has a sign error. I.e., $\delta q = q_p-q_e$ with $q_e$ being positive and with the actual electron charge equal to $-\!q_e$. (That is also consistent with $\delta q=q_n$ in case of $q_\nu=0$ as mentioned in \cite{hughes}.) Such a suggestion is consistent with their absolute value of $e_\nu$, but we arrive at a different sign. (The cited paper gives a positive central value but for $\overline{\nu}_e$.)
We do not use the values from \cite{hughes} in our averages (see below).

They \cite{hughes} also produced a constraint under an additional assumption of $e_\nu=0$.
As a verification of our understanding of the sign problems, we reproduce their $q_n$ under assumption of $q_\nu=0$, but do not use it further. Still, it is important to emphasize that they clearly recognized that those conditions (the charge conservation and $e_\nu=0$) are two separate ones and should not be combined by default. The constraints from \cite{hughes} are included neither in the neutron section of \cite{rpp2022} nor in the one that considers the proton-electron charge difference. The conservation of the electric charge (\ref{eq:beta:pennu}) that contains $e_\nu$ enable the experiments on the neutrality of the atomic and molecular particles to be applied to the neutrino charge. None of such constraints was considered as a contribution to the neutrino-charge section of \cite{rpp2022}; the constraints on $e_\nu$ are considered there in a separate section containing only those that obtained by directly constraining the neutrino charge, which automatically excludes the laboratory constraints on molecular and atomic particles and the neutron, given in Table~\ref{tab:q:data}.

The neutrino constraint, that dominates in \cite{rpp2022}, was obtained in \cite{caprini}
\begin{equation}\label{eq:const:cosm:nu}
|e_\nu|\leq4\times10^{-35}\,e\,~~~({\rm CL}=95\%)\,.
\end{equation}
It is very strong comparing with the level of laboratory constraints on the neutrality of subatomic, atomic, and molecular species given in Table~\ref{tab:q:data}. The $e_\nu$ constraint is derived from cosmological data. The other data overviewed in \cite{rpp2022} (see also \cite{das,review1,review2}) are not used for the PDG constraint being many orders of magnitude weaker than the one from \cite{caprini}. They also are essentially weaker than the level of the laboratory constraints discussed above.
While PDG ignores laboratory constraints on $e_\nu$ such as discussed above and similar, some reviews of neutrino properties consider them \cite{das,review1,review2}.

Constraint (\ref{eq:const:cosm:nu}) is based on the following assumptions, which we find controversial. At the first stage the overall spacial electric-charge density in the Universe is constraint and that is the next stage to interpret the spacial density in terms of the neutrino charge density that requires the disputed assumptions.
\begin{itemize}
\item The electric charge is supposed \cite{caprini} to conserve now, but not in a certain early epoch when the overall charge of the Universe may be generated. The mechanism of such an overall charge is not discussed, however, at the current epoch the overall charge density should be express in terms of a current particle density for each species and their current values of the charge. That allows for a constraint on the charge of the neutrino as it is now.
\item There may be several contributions to the charge density, but it is suggested in \cite{caprini}, there is no cancellation between them. To have the result on $e_\nu$ derived in \cite{caprini} literarily correct there should be no cancellations at all. To have the result correct by the order of magnitude there should be no strong cancellation, i.e., the individual electric-charge density of the neutrinos should be comparable with the overall one.
\item The charge density of neutrinos is estimated in \cite{caprini} through a product of a possible neutrino charge $e_\nu$ and the current neutrino number density
\begin{equation}\label{eq:nnu}
n_\nu=115.05\;{\rm cm}^{-3}\;.
\end{equation}
The figure is in fact the number density of neutrinos and antineutrinos of one kind within the standard cosmology, that suggests that 6 individual particle densities (3 kinds of $\nu$'s $\times$ (neutrinos or antineutrinos)) are approximately the same and equal to the half-value of (\ref{eq:nnu}) (see, e.g., \cite{peebles,weinberg_co,zuber}). (The standard [approximate] result on $n_\nu$ of different kinds of $\nu$'s and their relation to the number of CMB photons $n_\gamma$ ignores a small excess of baryons over antibaryons and possibly of leptons over antileptons.)

Terminology, while considering thermodynamics of [very] early Universe, may be somewhat confusing  (see, e.g., \cite{peebles,weinberg_co}). Usually, $n_\nu$ includes both neutrinos and antineutrinos as different degrees of freedom of the same particle and the mentioned standard value of $115\;{\rm cm}^{-3}$ is such a value. The prediction of the number of particles is possible because the number of particles follows any change of the temperature that, e.g., may be caused by the annihilation of heavy particles into the light ones or a phase transition of the vacuum. The [neutral] bosons such as photons may in principle be born by one, however, the fermions are born in pairs. There are several processes that are responsible for the energy transfer between different degrees of freedom, e.g., from electrons to neutrinos etc. Some of them, such as $e^+e^-\leftrightarrow \nu{\overline{\nu}}$ are also responsible for the number of neutrinos to be at the equilibrium. In other words, the standard cosmological model suggests that we always deal with [approximately] the same number of the neutrinos and antineutrinos.

According to \cite{caprini} there was an unspecified mechanism to generate overall electric charge in unspecified epoch. Applying the standard value for $n_\nu$ one suggests that such a mechanism played not more than a negligible role in the thermodynamic phenomena of the Universe when $n_\nu$ was formed and therefore the standard relation between the number of neutrinos and antineutrinos holds. The relation is the equality with negligible corrections. Besides, a {\em large\/} difference in the number of neutrinos and antineutrinos would suggest a {\em large\/} asymmetry between leptons and antileptons, which is unlikely, and a strong violation of the conservation of the leptonic number.

Therefore, the very use of (\ref{eq:nnu}) means the acceptance of the standard cosmological model. It is worth to mention that the latter is a self-consistent model that is confirmed by numerous observations (see, e.g., \cite{PLANCK}). Therefore as long as we speak about the numbers comparable with $n_\nu$ (and $n_\gamma$), we should observe a strong cancellation between neutrino's and antineutrino's contributions to the spacial charge in the Universe if they obey theoretical CPT invariance or phenomenological established relations between the charge of particles and antiparticles for many species under assumption of the charge conservation (at the present epoch). The alternative of such a cancellation is to suggest that there is a strong dominance of neutrino over the antineutrino (or {\em vice versa\/}) which would immediately invalidate the value $n_\nu=115.05\;{\rm cm}^{-3}$.

We remind that is not important what the particle charge {\em was\/} when a certain kind of particles was created, it is important what {\em is\/} their charge now. At the present epoch, the charge of neutrinos and antineutrinos, if any, is opposite ($e_{\overline\nu}=-e_\nu$), while their number density is [approximately] the same which makes unavoidable a cancellation between their contributions into the value of the overall charge. However, the cancellation between the number of the neutrinos and antineutrinos is rather to be approximate.

It is expected that there may be an excess of one over the other. It is expected that such an excess, that explicitly manifests the lepton-antilepton asymmetry, would have a number of particles comparable with the number of baryons and electrons at our epoch. Those were created during baryogenesis or leptogenesis. At least one component of such an excess in the neutrino sector is related not to the genesis by itself but to the subsequent primordial nuclear synthesis. The decay of all free neutrons by a certain moment, keeping the bound (as constituents of the deuterons) ones intact, would make $n_n$ and the excess of neutrino over antineutrino correlated.

While the mechanism of the baryon-antibaryon asymmetry is unclear as well as the involvement of the leptons, the correlation between $n_{\overline\nu}-n_\nu$ and $n_n, n_p+n_e$ due to the primordial nuclear synthesis is well established and experimentally verified (see, e.g., \cite{weinberg_co}). That means that there may plausibly be correlations between $n_{\overline\nu}-n_\nu$ with other number densities due to the baryogenesis or leptogenesis, and there definitely {\em are\/} correlations between $n_{\overline\nu}-n_\nu$ with other number densities, such as $n_n$ and $n_p+n_e$ due to the nuclear synthesis.

The current values of their charges of the involved particles are also correlated through the conservation of the charge in the current epoch. That makes possible further cancellations between the number of particles for different components of `excessive' baryonic and leptonic particles of different kinds.
\end{itemize}
Concluding, with the standard cosmological theory (see, e.g., \cite{weinberg_co}), base theoretical principles (such as the CPT theorem), the charge conservation (at the present epoch), and phenomenological knowledge about contemporary particles, the `no-cancellation' assumption does not seem fulfilled. The almost complete cancellation between contributions of neutrinos and antineutrinos is expected by many orders of magnitudes. Since the excessive density contributions (i.e., the differences between numbers of particles and antiparticles of a kind) are correlated, further cancellations between individual contributions of various kinds of particles are possible. The latter may set the overall charge to zero, even with the individual contributions (of neutrino-antineutrino, protons and electrons, neutrons) being non zero.
Because of the controversy of the assumptions in \cite{caprini} we reject their constraint.

The `non-cancelation' hypothesis is a very strong and non-trivial assumption that in fact requires an alternative cosmological model of early Universe, which would provide a very different contents of the particles in the Universe to avoid the approximate equality of $n_\nu$ and $n_{\overline{\nu}}$ in the first place and other subsequent correlations in the number density of various particles. The model should nevertheless be consistent with the observed data.

In the meantime the PDG constraint on $e_\nu$ in (\ref{eq:const:cosm:nu}) is based on the evaluation in \cite{caprini} alone (see \cite{rpp2022} for detail), which makes it invalid as well and requires more consideration.

The PDG constraints are considered for $e_p+e_e$, $e_n$, and $e_\nu$ independently in separate sections of \cite{rpp2022} and in particular they do not use any information on the neutrality of atomic particles and the neutron for their neutrino constraint. As we see above the laboratory phenomenological constraints on the neutrality of subatomic, atomic, and molecular species produce a constrain for $e_\nu$ as well. That follows from the charge conservation in the $\beta$ decay. The PDG constraint on $e_n$ (\ref{eq:en:rpp}) is the average \cite{rpp2022} of results from \cite{baumann,bressi}, while the constraint on $e_p+e_e$ is based on the result of \cite{bressi} alone.
We object the numerical values for the PDG interpretation \cite{rpp2022} of the result in \cite{bressi} (which as a matter of fact follows the interpretation in the original paper) (see above), but not the level at which $e_n$ and $e_p+e_e$ are constraint, namely, the constraints (in units of $e$) are at the level of a part in $10^{-21}$. This level is essentially weaker than the one in (\ref{eq:const:cosm:nu}), that we have just dismissed because of their controversial assumptions. Considering the value
\begin{equation}\label{eq:nu:pen}
e_\nu=e_p+e_e-e_n
\end{equation}
from the laboratory constraints, we are not interested in constraints on $e_\nu$ that are essentially weaker than the results at the level of $10^{-21}\;e$. Noting that all the direct constraints on $e_\nu$ listed in \cite{rpp2022} (see, also \cite{das,review1,review2}) are essentially below the level of interest, we ignore them.

That concludes our overview of the data, we are to apply. Now we are to consider phenomenological consequences of the laboratory constraints in Table~\ref{tab:q:data}. The key theoretical instrument is the conservation of the electric charge in the $\beta$ decay of the neutron in its complete form (\ref{eq:beta:pennu}).

A QED theory with a charged massless particle is a controversial one \cite{masslessQED}.
E.g., one has to encounter problems with the vacuum-polarization contribution and the pair production. That is not a problem that the theory is inconsistent by itself, but a problem that such a theory is not consistent with the Coulomb's $1/r^2$ long-range interaction and
photon propagation function
that are a part of our phenomenological picture.

Therefore for the massless neutrinos it is reasonable to suggest $e_\nu=0$. We may set $e_\nu=0$ despite of the presence of the neutrino mass. That leads us to the phenomenological constraint on $e_n$ as obtained by PDG (see (\ref{eq:lab:con})) if, following the mentioned paper, we base our constraint on two most accurate values. The complete set of the constraints is
\begin{eqnarray}\label{eq:phen:0}
e_p+e_e&=&(-0.2\pm0.8)\times10^{-21}\,e\nonumber\\
e_n&=&(-0.2\pm0.8)\times10^{-21}\,e\nonumber\\
e_{\nu}&=&0\,.
\end{eqnarray}

For other evaluations through out the paper similarly to (\ref{eq:phen:0}) we use for the average values two most accurate constraints, namely, on the neutrality of $n$ \cite{baumann} and SF$_6$ \cite{bressi}. We have narrowed our consideration for several reasons. Those two results strongly dominate and the contribution of the rest is marginal; the use of those two results makes our constraints comparable with the ones in \cite{rpp2022}, where the data of the same two experiments were used; the most accurate results of the rest of laboratory measurements on the neutrality are the ones on K and Cs \cite{hughes} where some details, such as the sign of $q_e$ and possible misprints, are unclear from their publication.

The outcome in (\ref{eq:phen:0}) is more or less consistent with PDG \cite{rpp2022} as concerning $e_p+e_e$ and $e_n$. PDG deals with the constraints on $e_p+e_e$ and $e_n$ suggesting $e_\nu=0$ and still considers a nonzero value of $e_\nu$ (see (\ref{eq:const:cosm:nu})), based on the evaluation in \cite{caprini}, that is at the level essentially below $10^{-35}\,e$ and therefore may be neglected while considering $e_p, e_n$, and $e_e$ at the level of $10^{-21}\,e$. However, the consistency is reached only if it is noted that the $e_\nu$ constraint is essentially stronger than that on $e_p+e_e$ and $e_n$, while the constraints on all three values are considered independently.
(We remind that the PDG's constraints on $e_p+e_e$, and $e_n$ are based on the following data. (i) $e_n$ is obtained from the average of the experiments on SF$_6$ and $n$, while (ii) $e_p+e_e$ is only from the experiment on SF$_6$. That makes the constraints in \cite{rpp2022} to be at the same level as ours in (\ref{eq:phen:0}) but somewhat different for $e_p+e_e$. Notably, it is suggested that $e_\nu=0$. However, the combination of the resulting PDG constraints $(e_p+e_e)-e_n$ is consistent with zero but not equal to it, because of a different choice of the data for the related constraints.)

\begin{figure}[thbp]
\begin{center}
\resizebox{1.0\columnwidth}{!}{\includegraphics{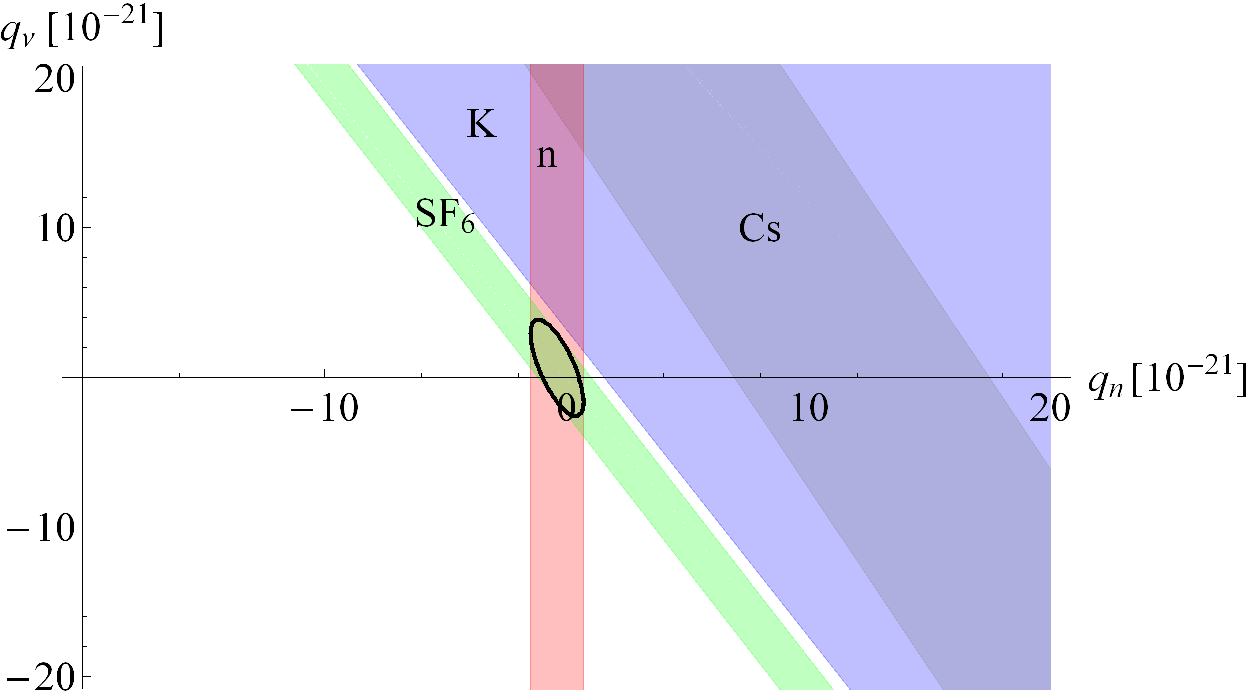}}
\end{center}
\vspace{-3.0mm}
\caption{Phenomenological constraint on the neutrino charge vs. the neutron charge from experiments on the neutrality of K and Cs \cite{hughes}, $n$ \cite{baumann}, and SF$_6$ \cite{bressi} (see Table~\ref{tab:q:data} for detail), based on the charge conservation in form of (\ref{eq:beta:pennu}). Our overall constraint, derived from the data on $n$ and SF$_6$ (see (\ref{eq:phen:m})), is represented by the standard ellipse. Here: $q_i=e_i/e$.}
\label{f:charge:nnu}       
\vspace{-4.0mm}
\end{figure}

In the meantime, since the neutrinos are massive, we have to consider a possibility for a non-zero value of $e_\nu$ from the beginning. With (\ref{eq:nu:pen}) we reformulate the individual constraints on the neutrality of atomic and molecular species (see Table~\ref{tab:q:data}) as (cf. \cite{das,review1})
\begin{equation}\label{eq:lab:con:nu}
A\,e_n + Z\,e_\nu = A\, R\,.
\end{equation}
The individual laboratory constraints are plotted in Fig.~\ref{f:charge:nnu}.

The results of combining data of two most accurate experiments \cite{baumann,bressi} are (cf. \cite{das})
\begin{eqnarray}\label{eq:phen:m}
e_p+e_e&=&(0.2\pm2.6)\times10^{-21}\,e\nonumber\\
e_n&=&(-0.4\pm1.1)\times10^{-21}\,e\nonumber\\
e_{\nu}&=&(0.6\pm3.2)\times10^{-21}\,e\,.
\end{eqnarray}
The results are correlated and are shown with the standard ellipse in the plots with the individual constraints in the $e_p+e_e, e_n$ and $e_\nu, e_n$ coordinates (see Figs.~\ref{f:charge:pen} and~\ref{f:charge:nnu}, respectively).

As mentioned, the Standard model is not a complete theory on all the physics that involves quarks, leptons and, in particular, neutrinos and as such does not set any constraint on the possible values of the electric charges of quarks and leptons. A complete theory would do that. One may produce somewhat stronger constraints, while considering certain extensions of the Standard model, as we do in the next section.

\section{Massive neutrino, the Standard Model, and related constraints\label{s:esm}}

The Standard Model of the electroweak and strong interactions (SM) is a model that suggests that $\nu_R$ does not exist and $m_\nu=0$. We {\em do\/} know that is incorrect. We may still consider a picture where $\nu_R$ does exist, but does not participate in any electroweak interactions, being present only in the mass term. That, in particular, requires that $e_\nu=0$ and therefore the right neutrinos do not contribute to the triangle anomaly. That would be very close to the SM, but such a suggestion is just a suggestion. (Because of the importance of the suggestion $e_\nu=0$ for the interpretation of the constraints in terms of the SM we consider the {\em a priori/} constraint $e_\nu=0$ and its theoretical consequences in detail in Sect.~\ref{s:enu}.)

The incomplete character of the SM has two consequences. One is that the SM Lagrangian is to be considered as a part of the complete one to which some neutrino details should be introduced not as corrections to the existing terms, but rather as additional terms. The other is that certain statements of the SM should be considered with a caution, since they are based on a consideration with an incomplete Lagrangian.

We are mostly interested in one of such statements. The SM describes the fundamental particles (the $u$ and $d$ quarks, the electron, and the electron's neutrino (only the left one) and their antiparticles as well as their heavier analogs from two other generations) as those that have specifically defined charges. The [assumed] values are not constraint at the so-called tree level, however the condition of the self-consistency of the theory includes a condition of the cancellation of the so-called triangle anomalies (see, e.g., \cite{weinberg}). If we consider the SM as a {\em complete\/} theory such a condition is a strong requirement. In terms of the fundamental particles the condition reads
\begin{eqnarray}
e_u&=&+\frac23\,e\,,\nonumber\\
e_d&=&-\frac13\,e\,,\nonumber\\
e_e&=&-e\,,
\end{eqnarray}
where we remind that $e$ is defined as $e=-e_e$ for any scenario, while in terms of the observable particles (hadrons rather than quarks) with $p=uud, n=udd$, the condition takes the form
\begin{eqnarray}\label{eq:particles:0}
e_p&=&+e\,,\nonumber\\
e_n&=&0\,,\nonumber\\
e_e&=&-e\,.
\end{eqnarray}
To above mentioned conditions we have to add the condition on the neutrino charge in the SM
\begin{equation}\label{eq:nu:0}
e_\nu=0\,,
\end{equation}
that follows from the absence of $\nu_R$ in the Lagrangian and therefore the absence of the mass term, which means a massless neutrino. (We mean here the Dirac-type mass term, since the Majorana mass is possible only within physics beyond the SM (BSM).)
The derivation of (\ref{eq:particles:0}) is described in detail, e.g., in Sect.~22.4 of \cite{weinberg}.

If we accept the above mentioned theoretical conditions, there is nothing to be constraint within the SM with $m_\nu=0$. The results on the values of interest
\begin{eqnarray}\label{eq:sm}
e_p+e_e&=&0\,,\nonumber\\
e_n&=&0\,,\nonumber\\
e_{\nu}&=&0
\end{eqnarray}
are trivial. No experimental data can change that within the SM. The conditions are also considered as the ones required for the quantization of the electric charge \cite{veltman,anomaly1,anomaly3,anomaly4}. A deviation from zero values above can in principle appears only from BSM new physics if there are new-physics contributions to the triangle anomaly. That in its turn would mean that the SM is incomplete also because of BSM contributions to electroweak phenomena. (Technically, we are not interested here in a complete theory in general, but only in a complete theory for the triangle anomalies.)

However, since {\em in vivo\/} $m_\nu\neq0$, the SM does not completely describe the reality even without any new BSM particles. Once we consider the Lagrangian of the SM as a {\em part\/} of the complete Lagrangian, we immediately realize that additional contributions to the triangle anomalies are possible (from $\nu_R$) and the constraints in (\ref{eq:sm}) are not valid anymore.

One may include $\nu_R$ and $m_\nu\neq0$ into the complete model in several ways and there is no consensus in that. Therefore we do not know how to unambiguously correct the mentioned conditions on the cancellation of the anomalies and should ignore them. That leads us to the phenomenological constraints with the assumption of the charge conservation (\ref{eq:beta:pennu}). Such a consideration and the derived constraints (see (\ref{eq:phen:m})) are consistent with the SM as a {\em part\/} of the complete theory with massive neutrinos. One may, e.g., consider a theory with no $\nu_R$ in the interaction part of in the Lagrangian, which would allow for the neutrino mass, but implies $e_\nu=0$ and will reproduce (\ref{eq:particles:0}) \cite{veltman,anomaly1,anomaly3,anomaly4}.

There is also a possibility of a `minimal extension' of the SM, assuming that we should introduce the neutrino masses without any additional violation of symmetries and without any additional field. That is possible if we treat the leptons the same way as the quarks (see, e.g., Sect.~22.4 of \cite{weinberg}), i.e., by introducing an ${\rm SU}_L(2)$ singlet $\nu_R^*$ additionally to $e_R^*$ in the lepton sector and treating the lepton pair $(\nu, e)$ the same way as the quark one $(u, d)$ (cf. \cite{veltman,jegerlehner}).
For the anomaly, such a model technically means that we keep the same number of the equations (as in the absence of $\nu_R$) as a condition of the cancellation of the anomalies, but add a new variable, the hypercharge of $\nu_R^*$. The equations without any $\nu_R^*$ contribution have only one valid solution that is consistent with the experiment. (The equations do not fix the coupling constant, so the solution is an expression of all the charges in units of one of them, say, $e=-e_e$.) If we add an additional term, let's parameterize it with $e_\nu$, we have for any [small] value of $e_\nu$ also one set of the results for $e_u$ and $e_d$ (in units of $e_e$ and new `free parameter' $e_\nu$) that is consistent with the experiment. Repeating the evaluation (e.g., the one in Sect.~22.4 with the neutrino's contributions added the same way as the electron's ones) we arrive at the constraints that allow for any small value $e_\nu\neq0$. Since the value of the latter is expected to be very small (as follows from experiments) we formulate the results in terms of the observable particles since only values of their charges can be measured with a high accuracy. The condition of the cancellation of the triangle anomalies reads
\begin{eqnarray}\label{eq:esm}
e_p+e_e&=&0\,,\nonumber\\
e_n+e_\nu&=&0\,.
\end{eqnarray}
The related condition in terms of the fundamental particles can be resolved as $e_u=+2/3\,e+1/3\,e_\nu,\; e_d=-1/3\,e-2/3\,e_\nu,\; e_e=-e$, while $e_p=e,\; e_n=-e_\nu$ (cf. \cite{review1,review2,das}).

\begin{figure}[thbp]
\begin{center}
\resizebox{0.95\columnwidth}{!}{\includegraphics{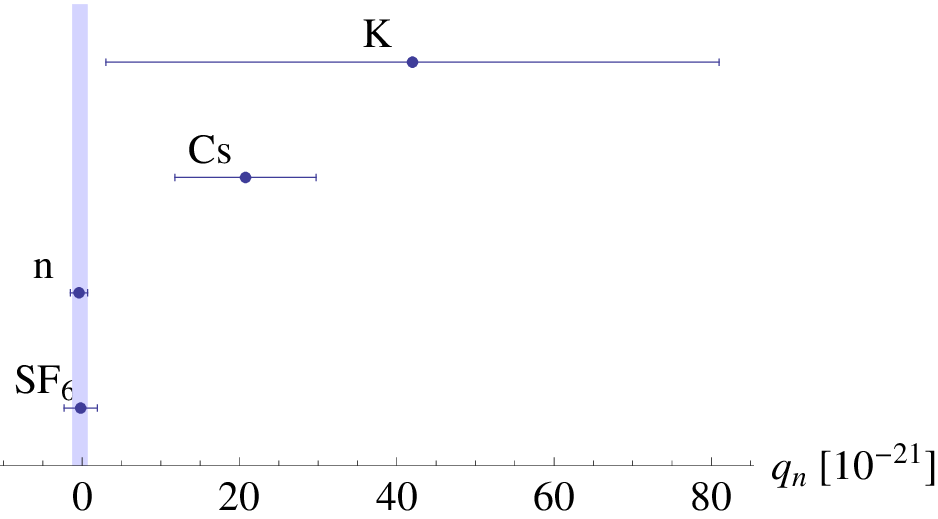}}
\end{center}
\vspace{-3.0mm}
\caption{Constraint on the neutron electric charge $e_n$ within the extended SM from experiments on the neutrality of K and Cs \cite{hughes}, $n$ \cite{baumann}, and SF$_6$ \cite{bressi}, based on (\ref{eq:esm}). Our average (see (\ref{eq:expsm})) over two the most accurate constraints (from the $n$ and SF$_6$ experiments) is represented by the vertical band. Within the extended SM $e_\nu=-e_n$. Here: $q_n=e_n/e$.
}
\label{f:charge:n}       
\vspace{-4.0mm}
\end{figure}

That means a narrowing the conditions for the evaluation of the individual data on the neutrality of atomic and molecular species (see Table~\ref{tab:q:data}). With $e_p+e_e=0$ the experimental bands (in the plots in Fig.~\ref{f:charge:pen} and Fig.~\ref{f:charge:nnu}) defined in (\ref{eq:lab:con}) and (\ref{eq:lab:con:nu}), correspondingly, shrink to data points
\[
e_n = \frac{A}{A-Z}\,R\;,
\]
that are plotted in Fig.~\ref{f:charge:n}. Note, our $e_n$ value from the SF$_6$ experiment \cite{bressi} is weaker than the one published in there because they set a constraint on the charge conservation as (\ref{eq:beta:pen}) that additionally of the charge conservation as (\ref{eq:lab:con:0}) sets also $e_\nu=0$. Their constraints is $e_n=R$, where the related $R$ value is given in Table~\ref{tab:q:data}. The interpretation of the atomic and molecular constraints (ours and their) is similar in terms of the number of variables, since they are both obtained by a relation on $e_p+e_e$. However, the relation in \cite{bressi} and
\cite{rpp2022} is $e_p+e_e=e_n$, while the relation within the extended SM is $e_p+e_e=0$. That does not affect the level of the constraint but does affect its numerical value. That also has different consequences for $e_\nu$. The constraint in \cite{bressi} and the neutron-charge section of \cite{rpp2022} combined with the charge conservation in the neutron's $\beta$ decay immediately sets $e_\nu=0$, while the extended Standard Model allows a non-vanishing value $e_\nu=-e_n$.

Evaluating the experimental data within (\ref{eq:esm}) and using only two most accurate results (namely, the ones on $n$ and SF$_6$) to provide the compatibility with the PDG evaluation \cite{rpp2022} we arrive at
\begin{eqnarray}\label{eq:expsm}
e_n&=&(-0.4\pm1.0)\times10^{-21}\,e\,,\nonumber\\
e_{\nu}&=&(0.4\pm1.0)\times10^{-21}\,e\,.
\end{eqnarray}
The theoretical constraint above is somewhat stronger than the phenomenological one in (\ref{eq:phen:m}) for $e_n$ and $e_\nu$, but at the same level as the latter; the constraint within the minimally extended SM in (\ref{eq:expsm}) assumes exact zeros for $e_p+e_e$ and $e_n+e_\nu$.

\section{On condition $e_\nu=0$ applied through some evaluations\label{s:enu}}

With the PDG constraints \cite{rpp2022} in the hands, where $e_\nu$ is constraint at level of $10^{-35}\,e$ from cosmological considerations \cite{caprini}, while $e_n$ and $e_e+e_p$ only at the level of $10^{-21}\,e$, the condition $e_\nu=0$ for the interpretation of the data on the neutrality of subatomic, atomic, and molecular objects \cite{baumann,hughes,bressi}, reviewed above in Sect.~\ref{s:pheno}, looks reasonable at least for an empiric evaluation. However, with the dismissal of the cosmological constraint on $e_\nu$ from \cite{caprini}
(see Sect.~\ref{s:pheno} for detail), the phenomenological situation has changed and the constraint on $e_\nu$ should also come from the same laboratory data (see Table~\ref{tab:q:data}) as the ones on $e_n$ and $e_e+e_p$ since the laboratory data are the most accurate one after the dismissal the above mentioned cosmological constraint.

There are different evaluations of such a kind of data in the literature and while some of them apply suggestion $e_\nu=0$ by default (see, e.g, \cite{bressi,rpp2022}) the other do not do so (see, e.g., \cite{hughes,das,review1}). We believe that is an important difference of the phenomenological and theoretical importance and we do not use the suggestion $e_\nu$ in this paper for the following reasons.

Considering the phenomenological constraints, the suggestion $e_\nu=0$ is unnecessary. As long as we have {\em two\/} constraints with different values of $A,Z$ and of a comparable strength (and we have them \cite{baumann,bressi}---see Table~\ref{tab:q:data} for details), the suggestion does not make the constraints on $e_p+e_e$ and $e_n$ much stronger, but makes them model-dependent. Meanwhile, the value of $e_\nu$ is assumed instead of to be constraint, which can be and should be done using the same data (see Sect.~\ref{s:pheno} for detail). Indeed, for the empiric constraints the model-independent approach is preferable. (Still such suggestions are often applied for the data evaluation when the authors prefer to obtain rather a model dependent constraint on their own than to combine their data with the one from other sources. In particular, the suggestion allows one to derive the constraints on $e_p+e_e$ and $e_n$ from the SF$_6$ data alone as it is done \cite{bressi}.)

Considering the consistency of the suggestion of $e_\nu=0$ and the SM (see Sect.~\ref{s:esm} for detail), we find that the suggestion together with the assumption of the cancellation of the triangle anomaly within the SM immediately leads to the trivial solution $e_p+e_e=e_n=0$, while the released value of $e_\nu$ leaves a more flexible parameter space and allows for a nontrivial solution $e_p+e_e=e_n+ e_\nu =0$, but $e_n=-e_\nu\neq0$. In other words, a search for a nontrivial value of the electric charge of the neutron $e_n$ is more natural and more consistent with the SM while suggesting $e_\nu\neq0$ rather than with $e_\nu=0$. It is worth to remind, that the SM is not a complete theory of the neutrino’s physics because of the absence of the neutrino mass term in the SM Lagrangian. It is possible that the additional neutrino terms in the complete Lagrangian produce contributions to the triangle anomaly as explained in Sect.~\ref{s:esm}. That makes the [small] deviation of the electric charges from their standard values possible. However, setting $e_\nu=0$, we do not allow the new neutrino contribution to the anomaly to appear unless we assume a nontrivial new physics, which makes the case of $e_\nu=0, e_n\neq0$ compatible with neither the SM nor with its principles as they are. We need to suggest something beyond the SM. It is much easier to adjust the SM to the case of $e_\nu\neq0, e_\nu+e_n=0$ than to the case of $e_\nu=0, e_n\neq0$ (see Sect.~\ref{s:esm} for detail).

It is helpful to compare condition $e_\nu=0$ with the condition of the conservation of the electric charge in the $\beta$ decay $e_n=e_p+e_e+e_{\overline{\nu}}$. The charge conservation is a cornerstone of the contemporary physical theory and a violation of it should lead to problems both in quantum and classical electrodynamics. That is a requirement of the consistency of the current theoretical framework. That is not a question of a theory of any particular phenomena, but about the base principle of the electrodynamics, namely, the gauge invariance. The theory cannot be `slightly’ modified in order to incorporate the non-conservation. The suggestion on the charge conservation is the suggestion of the consistency and validity of some key elements of the current theoretical framework, that are established experimentally.

On contrary, the suggestion on $e_\nu=0$ is not required for the consistency of the current theoretical principles and the theory as it is (i.e., the SM without $\nu_R$ + an empiric description of the neutrino socillations) and can be easily modified to accommodate a small value of $e_\nu$ (see SEct.~\ref{s:esm}). So, that is a suggestion additional to the current framework and its use means a certain model dependence and has consequences. As mentioned above about adjusting the SM to the case of $e_\nu\neq0, e_\nu+e_n=0$ and the case of $e_\nu=0, e_n\neq0$ we see that the suggestion $e_\nu=0$ is not harmless from the theoretical point of view. In other words, theoretically, taking into account the consequences, the suggestion $e_\nu=0, e_n\neq0$ is a very strong and nontrivial one.

Concluding the consideration of the zero neutrino-charge, $e_\nu=0$ is neither a good suggestion nor a natural one for the search of $e_n\neq0$. The question is not about which suggestion to take. No additional suggestions are required to evaluate the data on the neutrality of the molecules, atoms, neutrons once we have two pieces with a high accuracy. Theoretically, the suggestion is also not the best choice.

It is worth to mention various consideration within the framework of the quantization of the electric charge \cite{anomaly1,newenu,anomaly2} and the uniqueness of the SM \cite{veltman,jegerlehner}. In the both cases the condition $e_\nu=0$ is a necessary requirement, however, it is considered not as a requirement of the SM as it is, but as an additional one.

Due to all the above mentioned reasons we consider the suggestion $e_\nu=0$ as an unnecessary model-dependent one and strongly recommend not to use it for the evaluation of the constraints on the electric charge of $e_n$. Besides, because of the controversy in the cosmological constraint, the source of the most accurate data on the neutrino charge is the atomic data and setting $e_\nu=0$ instead of constraining $e_\nu$ is counterproductive.

\section{Conclusions\label{s:concl}}

Some time ago neutrinos were believed massless which rather excluded any possibility for a [small] electric charge for them from the theoretical and phenomenological point of view. It also made the SM in its original form a complete theory for neutrinos and their electroweak interactions. With the massiveness of the neutrino established, a small value of its electric charge becomes possible and should be constrained. Possibility of $e_\nu\neq0$ in its turn leads to corrections in the interpretation of certain non-neutrino experiments, since the deviation from the standard charges of various particles, such as $0, \pm e/3, \pm 2e/3,\pm e$, are related through the conservation of the electric charge in weak-interaction processes. The best of well-controlled constraints on $e_n, e_p+e_e, e_\nu$ can be obtained in laboratory experiments on the neutrality of would-be neutral objects such as subatomic, atomic, and molecular particles \cite{bressi,baumann,hughes} (see Table~\ref{tab:q:data} for detail). We discuss above two kinds of constraints that are in principle consistent with the Standard Model (as a part of the complete theory that in particular contains massive neutrinos and all the electroweak interactions both with and without neutrinos). They are summarized in Table~\ref{tab:e:const}. They all are (in the units of $e=e_p$) at the level of a part in $10^{21}$, with certain values being exactly zero within certain assumptions.

\begin{table}[thbp]
\begin{center}
\begin{tabular}{ccc}
\hline
Quantity  & Phenomenological & Constraint \\
  & constraint & within extended SM \\
\hline
$e_p+e_e$ & $(0.2\pm2.6)\times10^{-21}\,e$ & 0 \\
$e_n+e_\nu$ & $(0.2\pm2.6)\times10^{-21}\,e$ & 0 \\
$e_n$ & $(-0.4\pm1.1)\times10^{-21}\,e$ & $(-0.4\pm1.0)\times10^{-21}\,e$\\
$e_\nu$ & $(0.6\pm3.2)\times10^{-21}\,e$ & $(0.4\pm1.0)\times10^{-21}\,e$ \\
\hline
\end{tabular}
\end{center}
\caption{Phenomenological constraints (see Sect.~\ref{s:pheno}) and constraints within the extended SM (see Sect.~\ref{eq:esm}) obtained in this paper from the results of \cite{bressi,baumann} (see Table~\ref{tab:q:data}) in the units of $e=e_p$.}
\label{tab:e:const}       
\end{table}

The phenomenological constraints deal with the SM as a theory that describes a part of the complete Lagrangian, a part that does not contain $\nu_R$ and does not include the neutrino mass term. `Phenomenological' means that we recognize that such an inclusion of the mentioned terms must be done, but we hesitate to theorize how that can be done. We emphasis that we consider the SM as a model with arbitrary charges of $e_p, e_e, e_n$, and $e_\nu$ as far as it is allowed. A real limitation on the values of the charges would come either from an assumption on their exact values that is based on experiments with a limited accuracy or
from a condition of the cancellation of the triangle anomalies once we consider the SM Lagrangian as a complete one for the anomalies (see, e.g., \cite{weinberg}). If the additional terms such as the $\nu_R$'s ones, contribute there, the standard condition is invalid and all the effect of the SM on the choice of the values of the charges are reduced to the conservation of the charge in the $\beta$ decay of the neutron as given in (\ref{eq:beta:pennu}).

One can also consider the SM as an instruction how to include $\nu_R$ and to construct the mass term of the neutrino (cf. \cite{veltman,jegerlehner}). That can be done by the introduction of $\nu_R$ in the same way as in the consideration of $u_R, d_R$, and $e_R$, with the $\nu$'s mass term generated by the same Higgs vacuum average that is responsible for the quark and electron masses. Further consideration of the neutrinos in the same way as the other fundamental fermions would allow for their charge. That would set a stronger constraint on $e_p+e_e, e_n$, and $e_\nu$ (from the same experimental data) than the ones from just the charge conservation. The theoretical constraint, that follows from the cancellation of the triangle anomalies, reads as in (\ref{eq:esm}). We consider such a theoretical construction as a [minimal] extension of the SM that combines the SM terms of the Lagrangian and the effects due to $m_\nu\neq0$, that both should be included in the complete theory of the electroweak processes with leptons.

To obtain the constraints for $e_n$ and $e_p+e_e$ we have to check the original constraints and in particular to correct the result of \cite{bressi} due to their assumption of $e_p+e_e=e_n$ (cf. \cite{hughes}). Neither the correction nor the consideration of the extended SM change the level of the constraint, which is a part in $10^{21}$ (in units of $e$), but affects their numerical values. Some of them are changed approximately by a factor of 3 (cf. \cite{rpp2022}). (An exception is $e_p+e_e$ which is exactly equal to zero for the minimally extended SM.) As concerning $e_\nu$ we have found that the result that delivers the PDG constraint \cite{rpp2022} is controversial and should be excluded. Our constraints on $e_\nu$ are several orders of magnitude stronger than the other original constraints on $e_\nu$ considered in \cite{rpp2022}. That happens because certain constraints such as (\ref{eq:hughes}) obtained in \cite{hughes} have been ignored there. Our constraints are two orders of magnitude stronger than that of \cite{hughes}.

Concluding, one sees that small values of $e_n$ and $e_\nu$ and even of $e_p+e_e$ are to a certain extend consistent with the SM. Since the constraints are very strong we could unlikely detect them in any phenomena except of dedicated observations. E.g., one may wonder whether the Rydberg constant extracted from spectroscopy of the hydrogen atom differs from the one from the deuterium atom (see, e.g., \cite{codata2018,PRL,PRD} and references therein). (That is a kind of a direct test whether the charge of the proton is the same as of the deuteron, i.e., whether $e_n=0$.) The related uncertainties (in $R_\infty$) are about 10 orders of magnitude above the constraint on the fractional difference between the charges of the proton and the deuteron. The other atomic-spectroscopy experiments are less sensitive to the issue. That means that as far as it concerns the precision atomic spectroscopy one may continue to ignore a possible value of $e_n\neq0$. Other related tests, such as the determination of the fine structure constant by different methods (see, e.g., \cite{codata2018,rb2011} and references therein), that can be interpreted as a probe on the universality of the coupling constant for the electron-electron electromagnetic interaction and for the electron-proton one, are less accurate by two and more orders of magnitude in fractional units that the comparison of different experimental values of the Rydberg constant (cf. \cite{PRL,PRD}). One more access to small deviations from the quantization of the charge can be within the interpretation of the mass spectrometry experiments that deal with the charge-to-mass ratios of atomic and molecular ions (see, e.g., \cite{MPIK_d,FSU_pd} and references therein). Their fractional uncertainty is somewhat below the one of the Rydberg constant.

The only experiments, not dedicated for a search of the would-be zero charge, where the charge at the level of $10^{-21}\,e$ might be in principle important, are tests of the equivalence principle. From point of view of the accuracy they are at the level of interest, however, neither laboratory experiments (where the gravity between Earth and different probe bodies was studied \cite{EP1}) nor astronomical experiments (with the motion of the Earth-Moon system at the presence of the Sun gravity was investigated \cite{EP2}) cannot control the balance between electrons and protons, a small unbalance of which can produce a small residual charge even in case of $e_p+e_e=e_n=0$.

We remind that a gravitational interaction of, say, two hydrogen atoms is by a factor of about $10^{-36}$ weaker than the electromagnetic interaction between their electrons. The constraint on the proportionality of the gravitational mass to the inertial one is at the level of several parts in $10^{13}$ \cite{EP1,EP2}, which means that an additional electromagnetic force because of unbalance of the electric charge in two gravitationally interacting molecules is at the level of $10^{-24}\,e$ per a gravitating nucleon. The test of the equivalence principle constraints not the excessive charge by itself but its non-universality per a nucleon that depends on the ratio of proton and neutrons (i.e., the details of the chemical and isotopic composition) if the number of protons and electrons is assumed to be balanced. The dependence on the composition should reduce the strength of the constraint to a certain extend, however, the lack of control of the balance of charged particles, such as the electrons and protons, makes it completely invalid.

We emphasize that the possibility for small nonzero values of $e_n$ and $e_\nu$ is perfectly consistent with the SM (as an established part of the complete Lagrangian) and does not require any new BSM physics. On contrary, condition $e_n=0$ or $e_\nu=0$, that is sometimes adopted by default, requires an explanation that highly likely assumes a certain physics beyond the SM.
If the mentioned values are zero that would mean that there is an additional symmetry principle that sets the zeros at the tree level and maintains them through the perturbation theory. That may happen in case of various Grand unification theories, of a Majorana mass of neutrinos etc. In principle, the constraint $e_\nu=0$ may be set not only theoretically (by accepting a certain BSM model) but be reached experimentally. In particular, a successful search of neutrinoless double beta decay (see, e.g., \cite{zuber}) could prove that. We would say that the zero values of $e_\nu$ and $e_n$ are rather desired than required for the contemporary standard physical picture.

We note that the concept of the elementary electric charge $e=-e_e=e_p$ with $e_n=0$ accepted by default by a broad physical community is not well established as an exact one. It greatly simplifies the considerations of the electromagnetic phenomena with atomic and subatomic particles and quantum-electrodynamics phenomena in particular, but it is not required for the validity of QED and its application to numerous precision applications \cite{QED,codata2018,PRL,PRD}.
Condition $e=-e_e=e_p$ should be appropriate within the SM (as the complete theory, which is incorrect) and within the [minimal] extension of the SM, while condition $e_n=0$ is appropriate only in the former case. Unless we arrive at experimental evidences concerning the neutrino sector of the theory we cannot make a conclusive statement on how the SM must be extended to accommodate the massive neutrinos. Until then we are to consider conditions $e_p+e_e=0$ and $e_n=e_\nu=0$ as possible and may be desired, but neither established by the experiment nor required for the theory.

\section*{Acknowledgments}

The author is grateful to Robert Szafron, Vladimir Ivanov, Thomas Udem, and Vitaly Wirthl for valuable stimulating discussions.
The work was supported in part by DFG (Grant  \# KA 4645/1-2).

\end{document}